\documentclass[sigconf, nonacm=true]{acmart}

\usepackage{multirow}
\usepackage{hyperref}

\usepackage[utf8]{inputenc} 

\AtBeginDocument{%
  \providecommand\BibTeX{{%
    \normalfont B\kern-0.5em{\scshape i\kern-0.25em b}\kern-0.8em\TeX}}}

\setcopyright{acmcopyright}
\copyrightyear{2020}
\acmYear{2020}
\acmDOI{000000000}

\acmConference[KDD '20]{ACM SIGKDD International Conference on Knowledge Discovery \& Data Mining}{August 22--27, 2020}{San Diego, CA}
\acmBooktitle{ACM SIGKDD International Conference on Knowledge Discovery \& Data Mining, August 22--27, 2020, San Diego, CA}
\acmPrice{15.00}
\acmISBN{978-1-4503-XXXX-X/18/06}

\begin{document}

\title{High-Resolution Air Quality Prediction Using Low-Cost Sensors}

\author{Thibaut Cassard}
\affiliation{\institution{Plume Labs}}
\email{thibaut.cassard@plumelabs.com}

\author{Grégoire Jauvion}
\affiliation{\institution{Plume Labs}}
\email{gregoire.jauvion@plumelabs.com}

\author{David Lissmyr}
\affiliation{\institution{Plume Labs}}
\email{david.lissmyr@plumelabs.com}


\begin{abstract}

The use of low-cost sensors in air quality monitoring networks is still a much-debated topic among practitioners: they are much cheaper than traditional air quality monitoring stations set up by public authorities (a few hundred dollars compared to a few dozens of thousand dollars) at the cost of a lower accuracy and robustness. This paper presents a case study of using low-cost sensors measurements in an air quality prediction engine. The engine predicts jointly PM$_{2.5}$ and PM$_{10}$ (the particles whose diameters are below $2.5~\mu m$ and $10~\mu m$ respectively) concentrations in the United States at a very high resolution in the range of a few dozens of meters.

It is fed with the measurements provided by official air quality monitoring stations, the measurements provided by a network of low-cost sensors across the country, and traffic estimates. We show that the use of low-cost sensors' measurements improves the engine's accuracy very significantly. In particular, we derive a strong link between the density of low-cost sensors and the predictions' accuracy: the more low-cost sensors are in an area, the more accurate are the predictions. As an illustration, in areas with the highest density of low-cost sensors, the low-cost sensors' measurements bring a $25\%$ and $15\%$ improvement in PM$_{2.5}$ and PM$_{10}$ predictions' accuracy respectively. In cities with the most low-cost sensors like Los Angeles and San Francisco, this improvement in the predictions' accuracy is very clearly reflected in air quality maps.

An other strong conclusion is that in some areas with a high density of low-cost sensors, the engine performs better when fed with low-cost sensors' measurements only than when fed with official monitoring stations' measurements only: this suggests that an air quality monitoring network composed of low-cost sensors is effective in monitoring air quality. This is a very important result, as such a monitoring network is much cheaper to set up.

\end{abstract}




\maketitle

\section{Introduction}

Air pollution is one of the major public health concern. World Health Organization (WHO) estimates that more than 80\% of citizens living in urban environments where air quality is monitored are exposed to air quality levels that exceed WHO guideline limits. It also estimates that $4.2$ million deaths every year are linked to outdoor air pollution \cite{WHO_report} exposure.

Despite those alarming figures, very few citizens have access to information about the quality of the air they breathe. More and more public and private initiatives are being developed to close this gap and give to citizens the information they need to protect themselves from air pollution.

This is a particularly challenging topic because air quality varies a lot, both in time and in space. For example, a polluted air can become clean in a few hours after a heavy rain. Also, a crowded street can be much more polluted than a green park area a few hundred meters away \cite{AQ_review}.

One of the key difficulties when it comes to air quality modeling is the lack of data: it is believed that there are about $30$ thousands official air quality monitoring stations set up and maintained by public authorities worldwide. This is orders of magnitude below the number of stations needed given how much air quality varies in space. Also, there is as far as we know no comprehensive database providing live and historical measurements for those official monitoring stations, and building this database is a very time-consuming task.

The quantity of official air quality data varies a lot depending on the location: urban areas in developed countries are generally reasonably well monitored, with dozens of official monitoring stations in cities like London or Beijing. However, rural areas and poorer countries may be covered very sparsely given the high cost of building and maintaining an air quality monitoring network (the official monitoring stations set up by public authorities cost generally between $\$ 10$k and $\$ 50$k per station).

Over the last few years, various low-cost air quality sensors have been developed. They are generally orders of magnitude cheaper than traditional monitoring stations and cost a few hundreds of dollars. They are able to measure the main air pollutants (from particulate matter to gaseous pollutants like NO$_2$ and O$_3$) and are generally less robust and accurate than official monitoring stations. Nevertheless, they provide a very valuable source of data and their use to enhance existing air quality monitoring networks attracts more and more attention.

This paper presents an engine able to perform PM$_{2.5}$ and PM$_{10}$ predictions in the United States. It is fed with measurements provided by hundreds of official monitoring stations across the country, measurements provided by a network of more than $4000$ low-cost sensors, and traffic data. The prediction engine is similar to the one introduced in \cite{deepplume} and it is trained to predict the official monitoring stations' measurements given their higher accuracy than low-cost sensors' measurements.

The paper is organized as follows. We discuss earlier works in Section 2. Section 3 gives a detailed overview of the data sources used by the engine. Section 4 presents the architecture of the engine and details the model estimation process. Section 5 provides an evaluation of the prediction engine.

\section{Related work}

The problem of air quality prediction is much studied in the literature and is tackled through various angles. \cite{deep_learning_review} and \cite{bigdata_summary} present comprehensive reviews of air quality modeling using machine learning approaches.

Some papers focus on air quality temporal forecasts and aim at predicting pollutants' concentrations at air quality monitoring stations using the stations' historical measurements and meteorological features. Most of them are based on neural networks (generally RNN or LSTM architectures) and the forecast horizon varies from a few hours to $48$ hours. \cite{deepair}, \cite{china_deep_rnn} and \cite{deep_cnn_lstm} build forecasting models in China, \cite{deep_air_net} and \cite{india_lstm_rnn} build models for indian cities.

Other papers propose spatiotemporal modeling frameworks. \cite{conv_lstm}, \cite{conv_lstm_2} and \cite{plume_net} use Convolutional LSTM networks introduced in \cite{conv_lstm_theory} for precipitation forecasting. The networks are fed with monitoring stations' measurements and meteorological features to build 24 hours to 48 hours air quality forecasts. The spatial resolution in \cite{conv_lstm} is $0.1$ degree. \cite{airnet}, \cite{deep_air_learning} and \cite{deep_air_quality} use similar deep learning architectures to build air quality forecasts in China. \cite{airnet} covers all China with a $0.25$ degree resolution, \cite{deep_air_learning} covers Beijing with a $1$ kilometer resolution and \cite{spatial_temporal} focuses also on Beijing. \cite{deep_distributed} provides fine-grained forecasts in $300$ Chinese cities using a deep learning spatio-temporal architecture.

\cite{global_approach}, \cite{r_package_airpred} and \cite{deepplume} focus on spatial air quality predictions. \cite{global_approach} builds global PM2.5 predictions based on monitoring stations and satellite-based measurements with a $0.1$ degree resolution using statistical modeling. \cite{r_package_airpred} models PM2.5 in the US using monitoring stations' measurements, atmospheric models outputs and land-use datasets. \cite{deepplume} presents a prediction engine very similar to the one used in this paper.

\cite{low_cost_accuracy}, \cite{low_cost_accuracy_2} and \cite{low_cost_accuracy_3} analyze low-cost sensors' measurements and remind that their accuracy is a very important concern when using them in air quality monitoring systems. \cite{multimodal_encoding} introduces a neural network architecture to perform data fusion from heterogeneous sensors. \cite{sensors_network} presents an air quality monitoring network where the sensors' measurements are processed in a neural network along with temperature and humidity.

Finally, a few papers show how using a network of mobile low-cost sensors can improve air quality predictions' accuracy. \cite{low_cost_survey} surveys the landscape of low-cost sensors for air quality monitoring and concludes that there is not enough large-scale studies and large datasets to show the value-added of low-cost sensors: that is the problem we are tackling in this paper. \cite{aclima_mapping} shows the results of experiments in Mountain View and San Francisco, \cite{mobile_network} presents modeling approaches on an experimental setup in Lausanne, Switzerland, and \cite{china_maps} presents a case-study in Beijing using mobile sensors.

\section{Data sources}

This section details the data sources used by the prediction engine. We introduce the euclidean distance $\| l - l' \|$ between two locations $l$ and $l'$. We define also the exponential kernel $k_d(l, l') = \exp\left(-\frac{\| l - l' \|}{d}\right)$, where the distance $d$ is expressed in kilometers.

\subsection{Official monitoring stations' measurements}

The United States official air quality monitoring network is formed with thousands of monitoring stations providing measurements on a hourly basis for the main pollutants harming people's health. As the low-cost sensors used in this paper provide measurements for PM$_{2.5}$ and PM$_{10}$ only, we focus on PM$_{2.5}$ and PM$_{10}$ which are monitored by $974$ and $328$ official monitoring stations respectively.

It is worth noting that the official monitoring stations provide very accurate and reliable measurements of particulate matter concentrations \footnote{See https://www3.epa.gov/ttn/amtic/files/ambient/pm25/spec/drispec.pdf for detailed specifications.}. Thus, those measurements are considered as the ground truth data the engine is trained to predict.

We include in the datasets all measurements from January 1st, 2019 to December 31st, 2019. We have found that there are missing and erroneous values (generally abnormally high) coming from those monitoring stations. They can be encountered during station maintenance windows, during station failures or if issues arise during the publishing or collection of said data. While we are not able to determine the exact cause of such errors, it is important to detect them and define an appropriate treatment: missing values are discarded from the datasets, and erroneous values are detected using an outlier detection engine and then discarded.

\subsection{Low-cost sensors' measurements}

The low-cost sensors considered in this paper form a network of several thousand outdoor geolocated sensors providing PM$_{2.5}$ and PM$_{10}$ measurements every few minutes along with other variables such as humidity and temperature. We include in the datasets all measurements from January 1st, 2019 to December 31st, 2019.

At any location $l$ in the United States, we define the low-cost sensor density as the weighted sum of the number of sensors around, the weights being computed with an exponential kernel $k_d$:
$$ \text{SensorDensity}(l) = \sum_{i \in sensors} k_d(l, l_i) $$

, where $l_i$ is the location of sensor $i$. In the experiments presented in this paper, $d$ has been set to $10$ kilometers.

\begin{figure}[h]
  \centering
  \includegraphics[width=\linewidth]{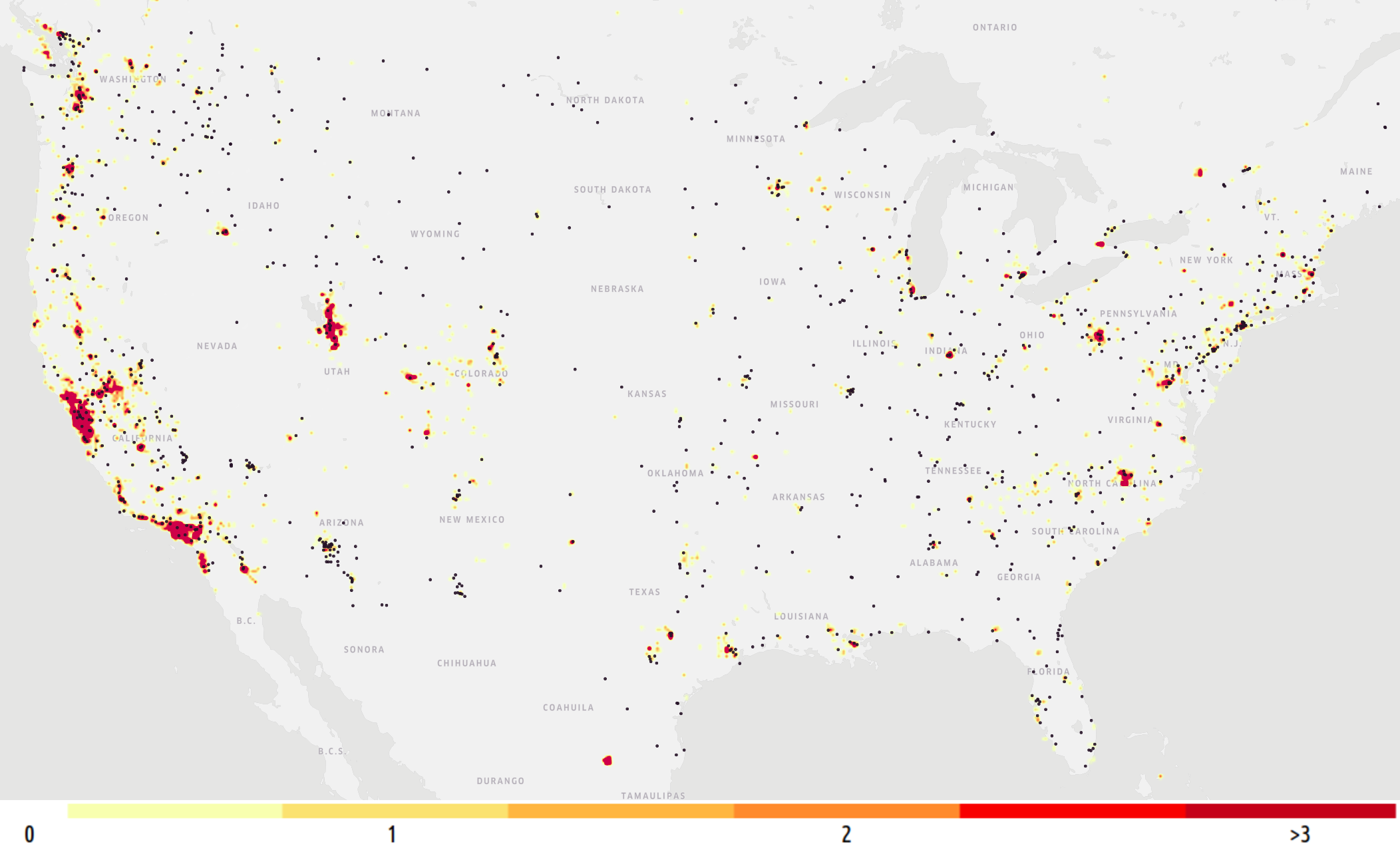}
  \caption{Official monitoring stations (black dots) and low-cost sensor density in the United States}
  \label{fig:stations}
  \Description{Official monitoring stations (black dots) and low-cost sensor density in the United States}
\end{figure}

Figure~\ref{fig:stations} shows the locations of the official monitoring stations and the low-cost sensor density in the United States.

\subsection{Road network and traffic data}

Those datasets are used as a proxy of vehicles' exhausts, which represent a significant part of air pollutants' emissions in cities.

\subsubsection{Road network}

We have road network details and topology in the regions covered by the prediction engine. Road network data is collected as a set of road segments, and any road segment is associated with a set of metadata regrouping a significant number of informations including a classification per usage, e.g., \textit{motorway} or \textit{residential}. We have built two aggregate categories named \text{Roads} and \text{MajorRoads}.

We define the features $\text{Roads}_d$ and $\text{MajorRoads}_d$ for every location $l$ as the weighted length of roads and major roads around location $l$, the weights being computed with an exponential kernel $k_d$. Those features estimate the road density around a location $l$, and can be interpreted as a rough proxy of the number of vehicles around.

\subsubsection{Traffic data}

We collect traffic data through a real-time \textit{jam factor} over each road segment of a given area. The \textit{jam factor} is a value between $0$ and $10$ measuring the road congestion: $0$ means that the road is not congested while $10$ means that it is very congested. Historical and real-time traffic data are collected across the United States.

For any distance $d$, we define the feature $\text{Traffic}_{t,d}$ at every location $l$ as the weighted sum of the jam factors of the road segments around location $l$, and each road segment is weighted by the product of its length and its functional class\footnote{The functional class is a classification of each road segment, from 1 (meaning a small road) to 5 (meaning a large road).}. The weights are computed with the exponential kernel $k_d$.

The product of the jam factor, the length and the functional class on a road segment is supposed to be proportional to the traffic emissions on this road segment. Hence, this feature can be interpreted as a proxy of traffic emissions around.

\subsection{Datasets' description}

We have built a dataset containing from January 1st, 2019 to December 31st, 2019 on a hourly basis the official monitoring stations' measurements, which form the ground truth data the prediction model is trained to predict, as well as the features used by the engine. Each data point corresponds to the measurements returned by an official air quality monitoring station at a given hour. The datasets have been built in Python using various packages including Scikit-Learn and Numpy.

The following features are built:
\begin{itemize}
    \item Features based on the $5$ closest official monitoring stations
    \begin{itemize}
        \item The measurements provided by the $5$ closest official monitoring stations $\text{Station}_{p,i}$, where $p \in [\text{PM}_{2.5} , \text{PM}_{10}]$ and $i \in [1,...,5]$
        \item The inverse distances to the $5$ closest official monitoring stations $\text{StationInvDistance}_{p,i}$
    \end{itemize}

    \item Features based on the $5$ closest low-cost sensors: we keep the last $16$ measurements of PM$_{2.5}$ and PM$_{10}$ concentrations, temperature and humidity
    \begin{itemize}
        \item The PM$_{2.5}$ and PM$_{10}$ measurements $\text{Sensor}_{p,t,i}$, where $p \in [PM_{2.5},PM_{10}]$, $t \in [1,...,16]$ and $i \in [1,...,5]$
        \item The temperature and humidity measurements $\text{Temperature}_{t,i}$ and $\text{Humidity}_{t,i}$, where $t \in [1,...,16]$ and $i \in [1,...,5]$
        \item The inverse distances to low-cost sensors $\text{SensorInvDistance}_{p,i}$, where $p \in [PM_{2.5},PM_{10}]$ and $i \in [1,...,5]$
    \end{itemize}

    \item Road network and traffic features $\text{Roads}_d$, $\text{MajorRoads}_d$ and $\text{Traffic}_d$. Several values have been considered for $d$, and it has finally been set to $100$ meters
\end{itemize}

It is worth noting that for each data point, the corresponding official monitoring station is excluded from the $5$ closest official monitoring stations, to make sure that the measurements the engine is trained to predict are not included in the features. Missing measurements are flagged as $NA$ and are treated specifically: it happens when a monitoring station monitors only one of the two pollutants predicted by our model.

\section{Models and estimation}

\subsection{Architecture of the prediction engine}

The prediction engine maps the features with the 2-dimensional vector giving the concentrations of PM$_{2.5}$ and PM$_{10}$ in $\mu g/m^3$. The engine is based on a somewhat classical neural network architecture with one hidden layer of $n_1$ units. The official monitoring stations' measurements, the road network features, the traffic estimates as well as the inverse distances to the official monitoring stations and low-cost sensors are provided as raw inputs to the engine.

We use a more advanced architecture to process the low-cost sensors' measurements: the last $16$ PM$_{2.5}$, PM$_{10}$, temperature and humidity measurements are processed using a one-dimensional convolutional neural network. This network is formed with two sequences of convolution and max pooling layers. Kernel size and number of filters are denoted $k_1$, $f_1$ and $k_2$, $f_2$ for the first and second one-dimensional convolution layers respectively, and pooling size is set to $2$ in both max pooling layers. The output of the convolutional neural network is then concatenated with the other features. It worked much better in practice than providing the raw low-cost sensors' measurements for the two following reasons:
\begin{itemize}
  \item A naive choice would have been to average the last measurements provided in the last hour. However, the measurements can vary significantly within an hour, in particular when a pollution peak happens, and processing them with a convolutional neural network enables to keep much more information
  \item Temperature and humidity are known to impact low-cost sensors' measurements accuracy. Low-cost sensors classify particle sizes by means of laser measurement. However, humidity can play a disturbing role during this measurement by increasing the size of a particle that adsorbs water or by creating droplets through condensation. This phenomenon is difficult to model and  depends on particle history. Hence, using them as features along with the pollutant concentrations measurements improved the prediction accuracy significantly
\end{itemize}

We consider $3$ different prediction models using different sets of features (see Table~\ref{tab:features}):
\begin{itemize}
    \item \textit{Station model}: this model uses official monitoring stations only and is used to assess how the official air quality monitoring network performs to predict air quality
    \item \textit{Sensor model}: this model uses low-cost sensors only and is used to assess how the low-cost sensors network performs to predict air quality
    \item \textit{Station and sensor model}: this model uses both official monitoring stations and low-cost sensors
\end{itemize}

\begin{table*}
  \caption{Features used in the prediction models}
  \label{tab:features}
  \begin{tabular}{c||c|c}
    \toprule
    Model & Air quality features & Other Features \\
    \midrule
    Station model & $\text{Station}_{p,i}$ & $\text{StationInvDistance}_{p,i}$, $\text{Roads}_d$, $\text{MajorRoads}_d$, $\text{Traffic}_d$ \\[5pt]
    Sensor model & $\text{Sensor}_{p,t,i}$ & $\text{SensorInvDistance}_{p,i}$, $\text{Temperature}_{t,i}$, $\text{Humidity}_{t,i}$, $\text{Roads}_d$, $\text{MajorRoads}_d$, $\text{Traffic}_d$\\[5pt]
    \multirow{2}{3.5cm}{Station and sensor model} & \multirow{2}{3cm}{$\text{Station}_{p,i}$, $\text{Sensor}_{p,t,i}$} & $\text{StationInvDistance}_{p,i}$, $\text{SensorInvDistance}_{p,i}$, $\text{Temperature}_{t,i}$, $\text{Humidity}_{t,i}$, \\
    & & $\text{Roads}_d$, $\text{MajorRoads}_d$, $\text{Traffic}_d$ \\[5pt]
  \bottomrule
\end{tabular}
\end{table*}

\begin{figure}[h]
  \centering
  \includegraphics[width=\linewidth]{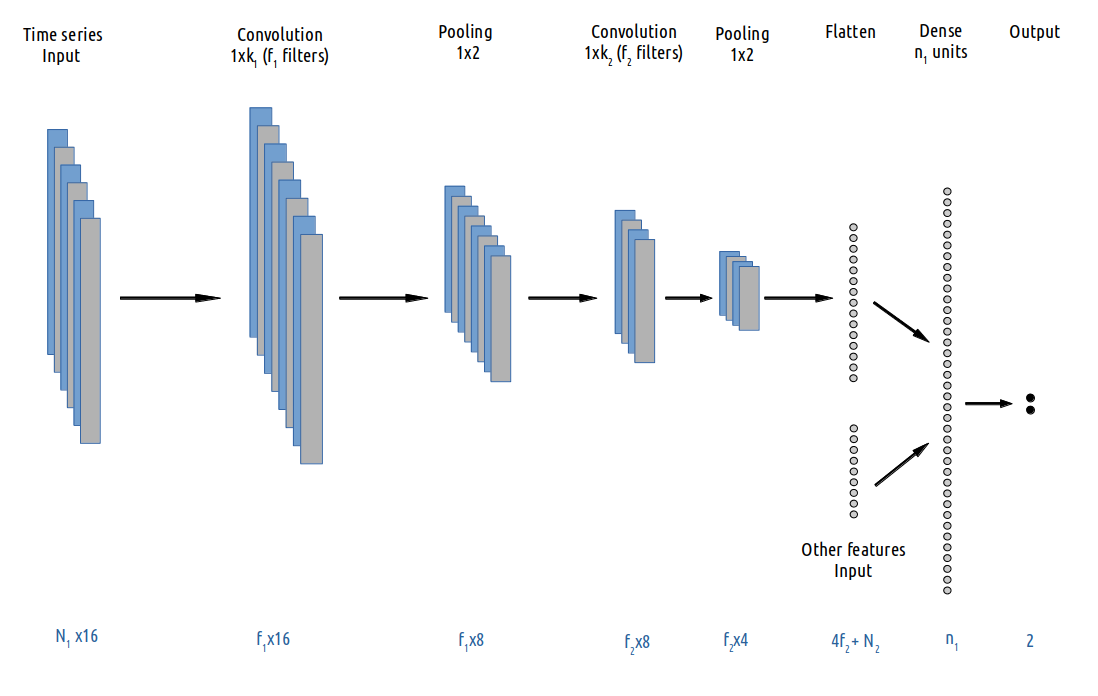}
  \caption{Models architecture}
  \label{fig:architecture}
  \Description{Models architecture}
\end{figure}

Figure~\ref{fig:architecture} summarizes the architecture of the models. $N_1$ and $N_2$ denote the number of features built from the low-cost sensors (PM$_{2.5}$, PM$_{10}$, temperature and humidity measurements) and the number of other features (official monitoring stations' measurements, road network and traffic data, inverse distances to the official monitoring stations and low-cost sensors) respectively. As the \textit{Station model} does not use low-cost sensors' measurements as an input, it is simply a fully connected neural network with a single hiden layer of size $n_1$.

\subsection{Estimation setup}

The set of official monitoring stations is split into two parts: $80$\% of the stations form the training set and the remaining $20$\% form the evaluation dataset. Given the key importance of the density of low-cost sensors in the analysis performed hereafter, the sampling of the official monitoring stations is performed within each decile of low-cost sensor density. This ensures that the distribution of low-cost sensor density remains similar in the training and evaluation datasets. Table~\ref{tab:number} shows the number of data points in the training and evaluation datasets.

\begin{table}
  \caption{Number of data points in the training and evaluation datasets (in millions)}
  \label{tab:number}
  \begin{tabular}{c|ccc}
    \toprule
     & Nb points & With PM$_{2.5}$ & With PM$_{10}$ \\
    \midrule
    Training set & 6.35 & 5.44 & 1.93 \\
    Evaluation set & 1.55 & 1.39 & 4.53 \\
  \bottomrule
\end{tabular}
\end{table}

The models are trained on the training dataset using TensorFlow and Keras. The loss used in training is the mean squared logarithmic error (MSLE) loss. $NA$ values are excluded from the loss computation. Hyperparameters of the models have been optimized and are as follows: $n_1 = 128$ in the $3$ models. In \textit{Sensor model} and \textit{Station and sensor model}, we use $k_1 = k_2 = 3$, $f_1 = 32$ and $f_2 = 8$. We use Adam optimizer with a learning rate equal to $0.001$. The models are estimated on $2$ epochs with mini batches of size $64$.

\section{Models evaluation}

This section evaluates the accuracy of the predictions built with the $3$ models introduced in the last section. The models are also compared to a simple benchmark predictor which consists in predicting for each pollutant the measurement provided by the closest official monitoring station.

\subsection{Prediction accuracy}

Table~\ref{tab:accuracy} gives the MSLE (loss used in training) and mean absolute error (MAE) computed on the evaluation dataset for the $3$ prediction models and the benchmark predictor.

\begin{table}
  \caption{MSLE and MAE computed on the evaluation dataset}
  \label{tab:accuracy}
  \begin{tabular}{cccc}
    \toprule
    Pollutant & Model & MSLE & MAE \\
    
    \midrule

    \multirow{4}{1cm}{PM$_{2.5}$} & Benchmark & 0.3327 & 3.4221 \\
                      & Station model & 0.2159 & 2.7535 \\
                      & Sensor model & 0.2424 & 2.9444 \\
                      & Station and sensor model & 0.2049 & 2.6419 \\

    \midrule

    \multirow{4}{1cm}{PM$_{10}$} & Benchmark & 0.4518 & 10.5989 \\
                      & Station model & 0.2910 & 8.2593 \\
                      & Sensor model & 0.3864 & 9.7116 \\
                      & Station and sensor model & 0.2860 & 8.2246 \\

  \bottomrule
\end{tabular}
\end{table}

We see that for PM$_{2.5}$ and for PM$_{10}$ and for both metrics computed (MSLE and MAE), the \textit{Station and sensor model} performs slightly better than the \textit{Station model}. The improvement in the evaluation MSLE is limited (about $5\%$ for PM$_{2.5}$ and $2\%$ for PM$_{10}$): the main reason is that most official monitoring stations across the United States have very few low-cost sensors around, hence the need to compare the prediction models in areas with a dense low-cost sensors network.

\subsection{Influence of the low-cost sensor density on the prediction accuracy}

\begin{figure}[h]
  \centering
  \includegraphics[width=\linewidth]{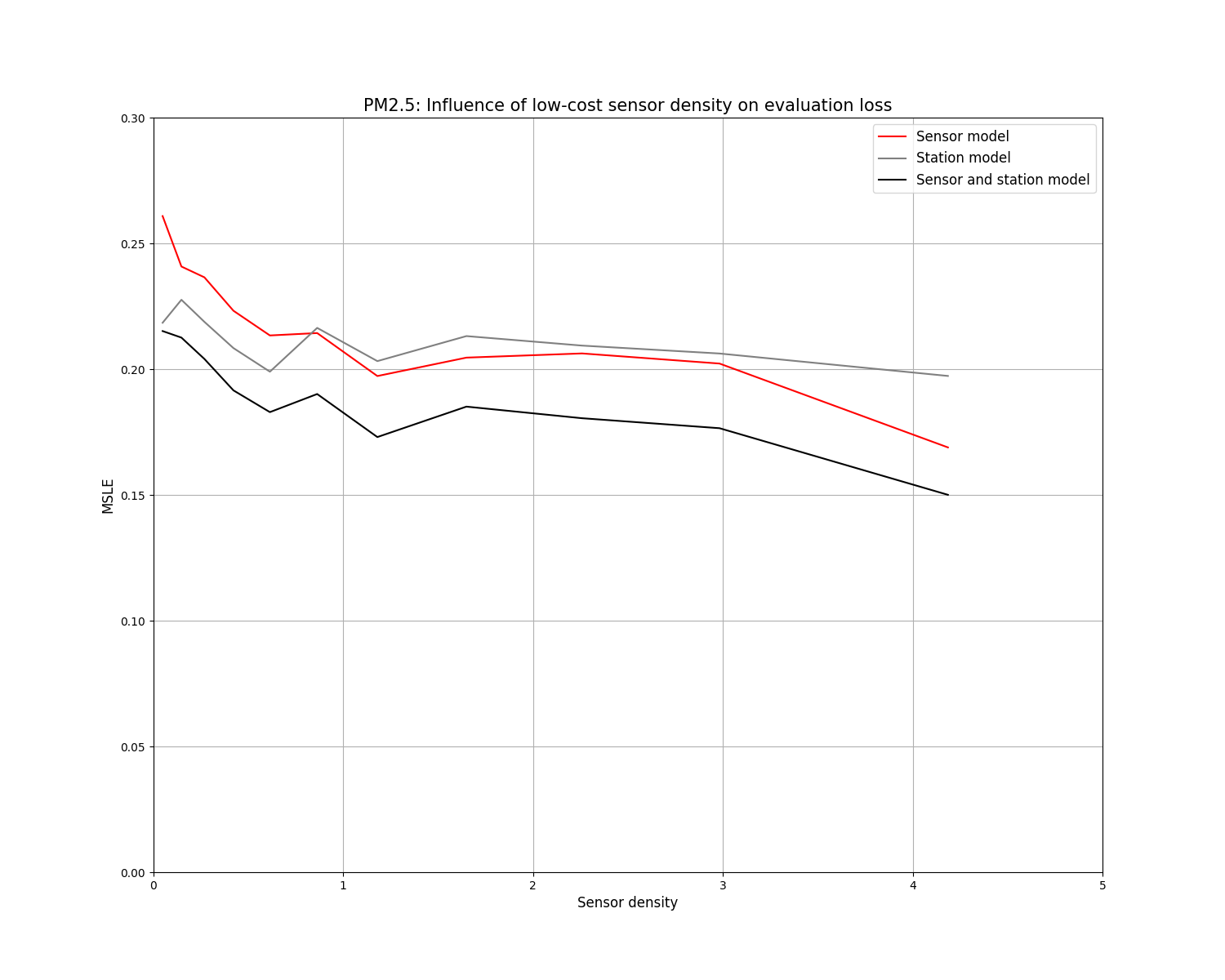}
  \caption{PM$_{2.5}$: Influence of low-cost sensor density on the evaluation loss}
  \label{fig:sensor_density_pm2_5}
  \Description{PM$_{2.5}$: Influence of low-cost sensor density on the evaluation loss}
\end{figure}

\begin{figure}[h]
  \centering
  \includegraphics[width=\linewidth]{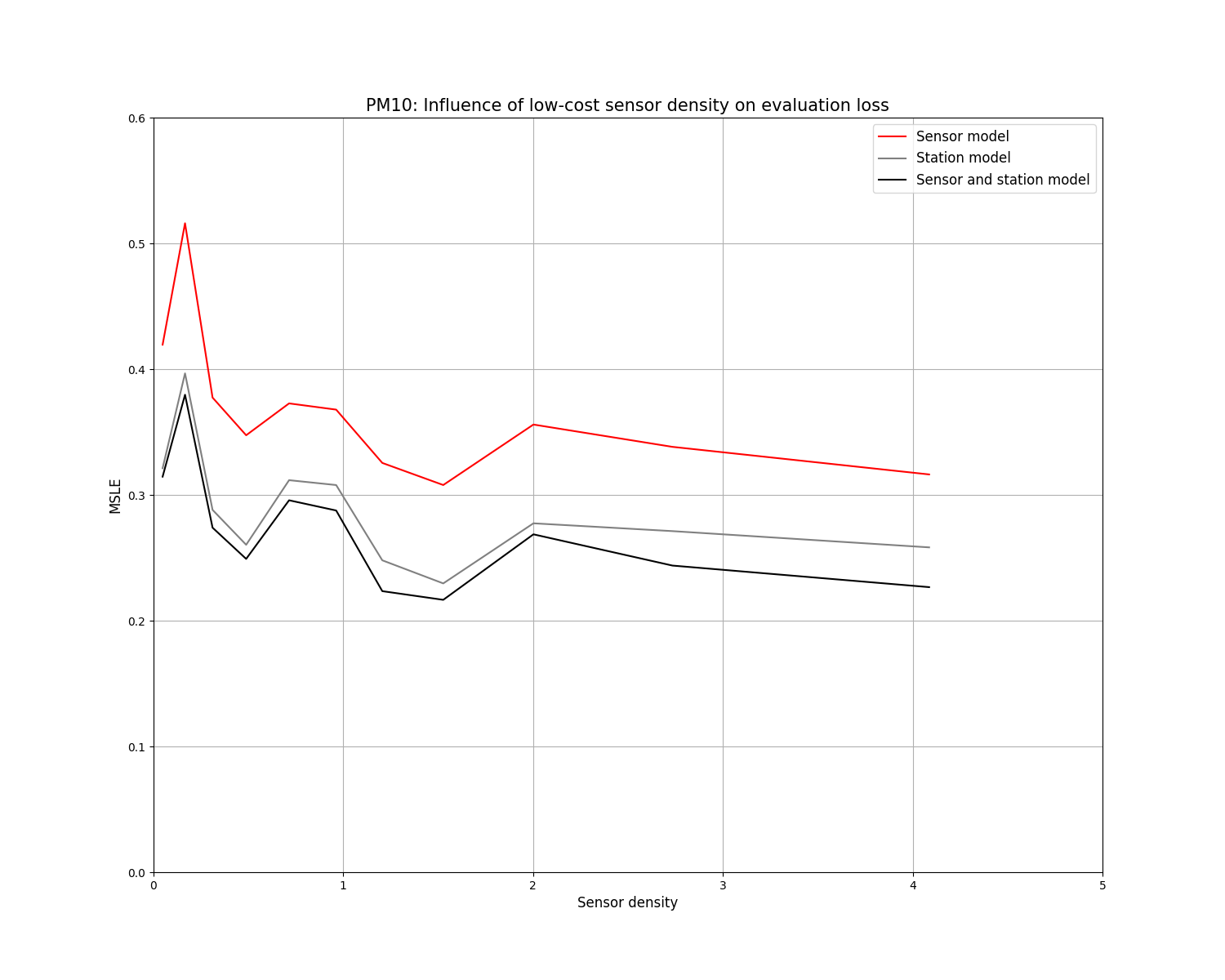}
  \caption{PM$_{10}$: Influence of low-cost sensor density on the evaluation loss}
  \label{fig:sensor_density_pm10}
  \Description{PM$_{10}$: Influence of low-cost sensor density on the evaluation loss}
\end{figure}

In this section, we remove from the datasets all data points whose low-cost sensor density is below the median value. Then, for PM$_{2.5}$ and PM$_{10}$, the datasets are split into $10$ batches based on low-cost sensor density: the batches contain about $400$ thousand and $150$ thousand data points for PM$_{2.5}$ and PM$_{10}$ respectively. Figures~\ref{fig:sensor_density_pm2_5} and ~\ref{fig:sensor_density_pm10} show the MSLE computed for the $3$ prediction models on each batch, for PM$_{2.5}$ and PM$_{10}$ respectively. The x-axis is the mean density on each batch.

The main conclusion is that the improvement in accuracy of the \textit{Station and sensor model} compared to the \textit{Station model} increases with the low-cost sensor density. In the batch with the highest low-cost sensor density (higher than $4$ in average), the improvements are $25\%$ and $15\%$ for PM$_{2.5}$ and PM$_{10}$ respectively. The improvement brought by the low-cost sensors is higher for PM$_{2.5}$: this is due to the greater accuracy of the low-cost sensors PM$_{2.5}$ measurements.

An other interesting conclusion is that for PM$_{2.5}$, \textit{Sensor model} is more accurate than \textit{Station model} as soon as the low-cost sensor density exceeds $1$.

\subsection{Focus on particular cities}

In this section, we focus on the $5$ cities with the highest number of low-cost sensors (hundreds of sensors in each). Table~\ref{tab:city_accuracy} compares the $3$ prediction models in those cities. The table gives also the average low-cost sensor density at the official monitoring stations which are the locations where the models are trained. The results for PM$_{10}$ are not given in San Francisco and Seattle because there are no PM$_{10}$ official monitoring stations in those cities.

\begin{table*}
  \caption{MSLE in particular American cities}
  \label{tab:city_accuracy}
  \begin{tabular}{cccccc}
    \toprule
    City & Pollutant & Mean sensor density around stations & Model & MSLE \\
    
    \midrule

    \multirow{6}{1.5cm}{Los Angeles} & \multirow{3}{1cm}{PM$_{2.5}$} & \multirow{3}{1cm}{2.70} & Station model & 0.2442 \\
                      & & & Sensor model & 0.1937 \\
                      & & & Station and sensor model & 0.1832 \\
                      & \multirow{3}{1cm}{PM$_{10}$} & \multirow{3}{1cm}{2.67} & Station model & 0.2388 \\
                      & & & Sensor model & 0.2456 \\
                      & & & Station and sensor model & 0.2240 \\

    \midrule

    \multirow{3}{1.5cm}{San Francisco} & \multirow{3}{1cm}{PM$_{2.5}$} & \multirow{3}{1cm}{3.73} & Station model & 0.1800 \\
                      & & & Sensor model & 0.1730 \\
                      & & & Station and sensor model & 0.1573 \\
    
    \midrule

    \multirow{6}{1.5cm}{Sacramento} & \multirow{3}{1cm}{PM$_{2.5}$} & \multirow{3}{1cm}{3.75} & Station model & 0.1925 \\
                      & & & Sensor model & 0.1523 \\
                      & & & Station and sensor model & 0.1493 \\
                      & \multirow{3}{1cm}{PM$_{10}$} & \multirow{3}{1cm}{4.51} & Station model & 0.1751 \\
                      & & & Sensor model & 0.2456 \\
                      & & & Station and sensor model & 0.1512 \\
    
    \midrule

    \multirow{3}{1.5cm}{Seattle} & \multirow{3}{1cm}{PM$_{2.5}$} & \multirow{3}{1cm}{2.78} & Station model & 0.1688 \\
                      & & & Sensor model & 0.1350 \\
                      & & & Station and sensor model & 0.1378 \\
    \midrule

    \multirow{6}{1.5cm}{Salt Lake City} & \multirow{3}{1cm}{PM$_{2.5}$} & \multirow{3}{1cm}{1.24} & Station model & 0.2155 \\
                      & & & Sensor model & 0.2370 \\
                      & & & Station and sensor model & 0.2030 \\
                      & \multirow{3}{1cm}{PM$_{10}$} & \multirow{3}{1cm}{1.28} & Station model & 0.3730 \\
                      & & & Sensor model & 0.4594 \\
                      & & & Station and sensor model & 0.3590 \\

  \bottomrule
\end{tabular}
\end{table*}

In all cities, the best performing model is \textit{Station and sensor model}. The improvement compared to \textit{Station model} is generally higher for PM$_{2.5}$ than for PM$_{10}$. In cities where the mean sensor density is higher than $2$, the improvement is very significant for PM$_{2.5}$: $25\%$ in Los Angeles, $13\%$ in San Francisco, $23\%$ in Sacramento and $18\%$ in Seattle.

\begin{figure}[h]
  \centering
  \includegraphics[width=\linewidth]{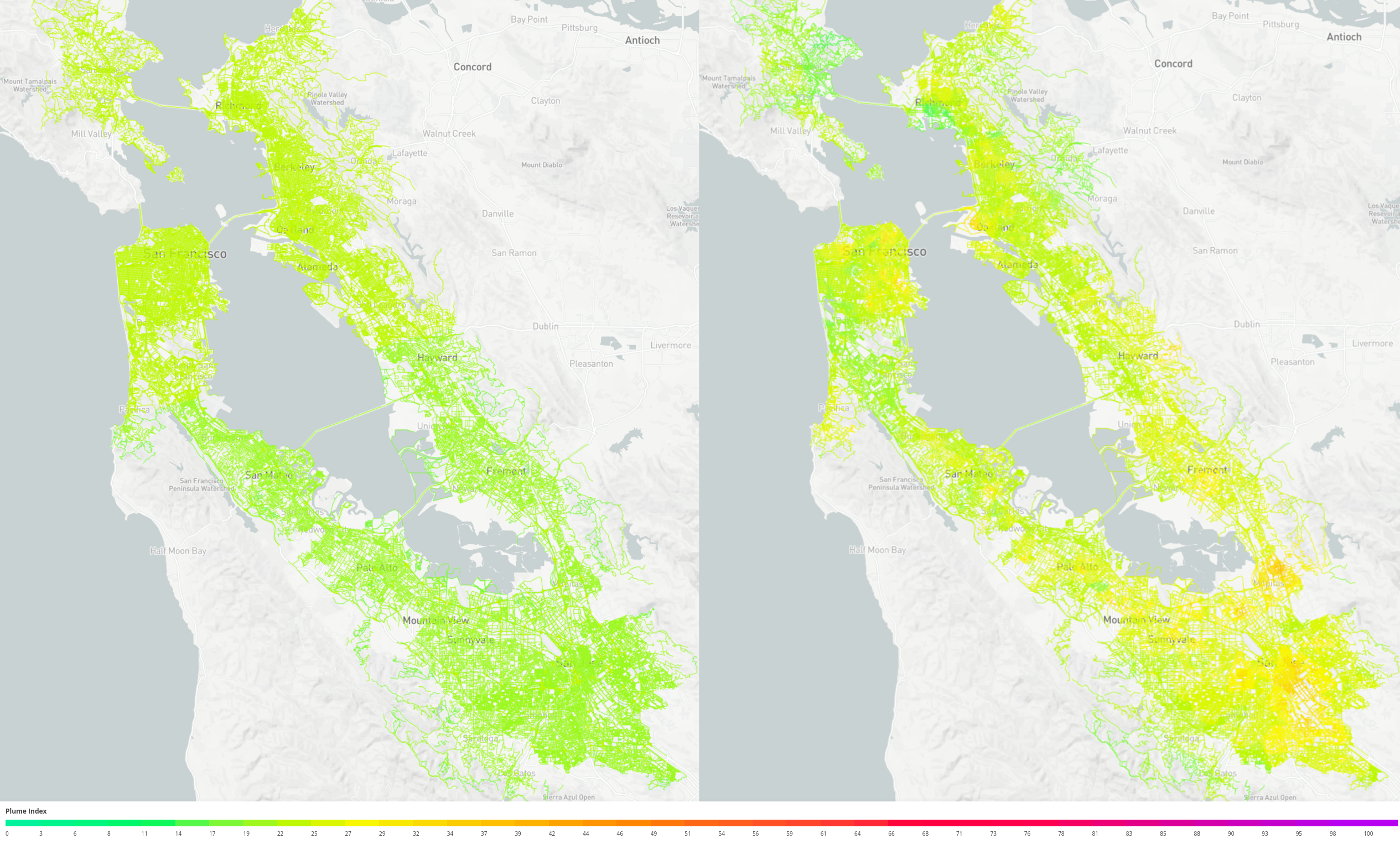}
  \caption{PM$_{2.5}$ map in San Francisco without and with using low-cost sensors' measurements (left and right respectively)}
  \label{fig:map_pm25}
  \Description{PM$_{2.5}$ map in San Francisco without and with using low-cost sensors' measurements (left and right respectively)}
\end{figure}

\begin{figure}[h]
  \centering
  \includegraphics[width=\linewidth]{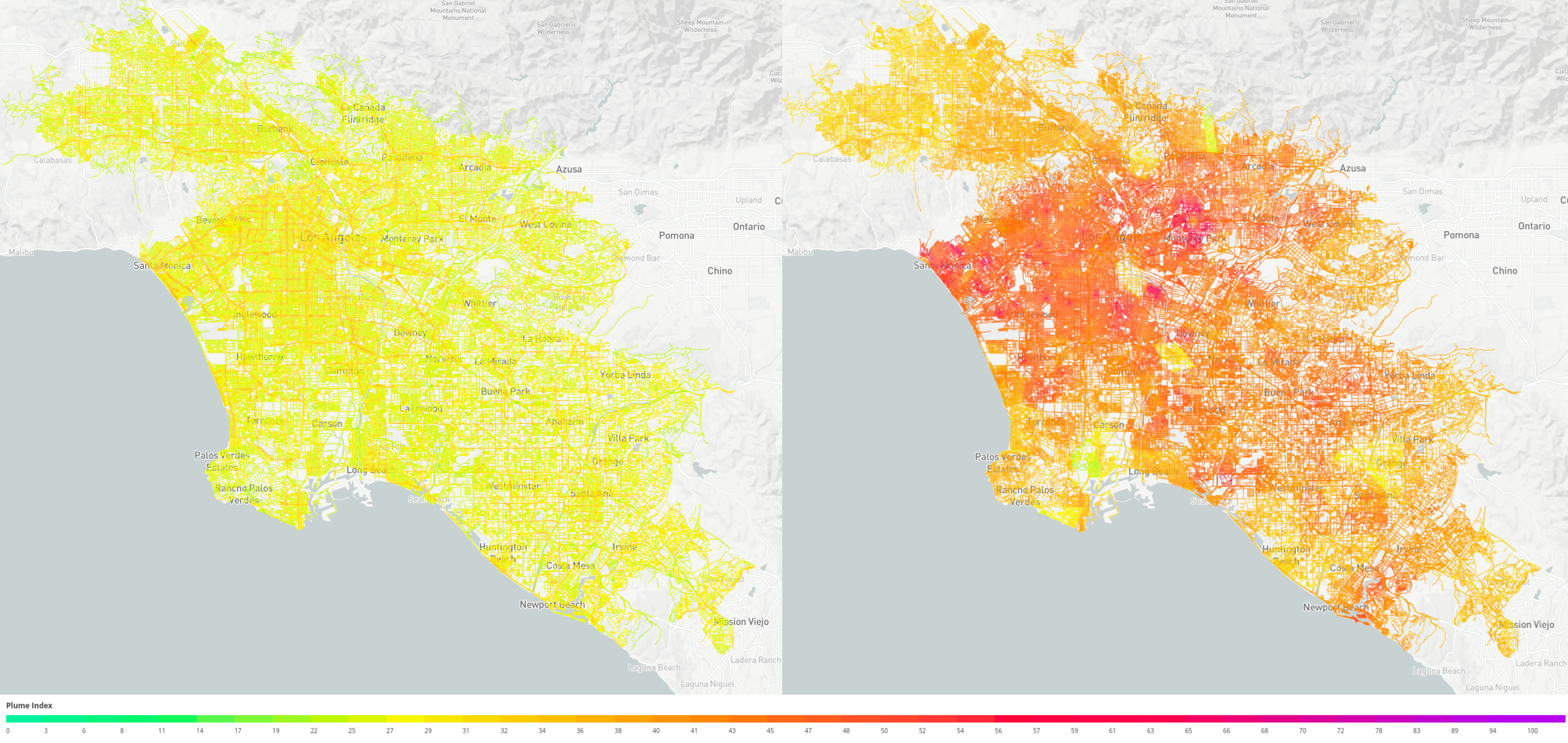}
  \caption{PM$_{10}$ map in Los Angeles without and with using low-cost sensors' measurements (left and right respectively)}
  \label{fig:map_pm10}
  \Description{PM$_{10}$ map in Los Angeles without and with using low-cost sensors' measurements (left and right respectively)}
\end{figure}

As an illustration, Figures~\ref{fig:map_pm25} and ~\ref{fig:map_pm10} show maps of PM$_{2.5}$ in San Francisco and PM$_{10}$ in Los Angeles realized at a given time. The left map is built with \textit{Station model} while the right map is built with \textit{Station and sensor model}. We see that the right maps show much more spatial variability.

\section{Conclusion and future work}

This case study of using low-cost sensors' measurements in air quality prediction is at our knowledge the only one produced at such a large scale, covering the whole United States and using the measurements provided by thousands of official monitoring stations and low-cost sensors. Thus, it enables to derive robust and very meaningful conclusions.

As expected, we noticed that low-cost sensors' measurements are less accurate and robust than official monitoring stations' measurements. However, we show that when part of a large network, low-cost sensors can be used to improve the prediction accuracy significantly. In particular, we have derived a strong link between the low-cost sensor density and the prediction accuracy.

An other conclusion is that in areas with a high density of low-cost sensors, a prediction model using low-cost sensors' measurements only performs better than a prediction model using official monitoring stations only: this result suggests that an air quality monitoring network composed of low-cost sensors is effective in monitoring air quality. This is important considering the price difference between traditional monitoring stations (several dozens of thousand dollars) and low-cost sensors (several hundred dollars).

The results presented here are obtained by assuming that the ground truth measurements the prediction model is trained to predict are the official monitoring stations' measurements. This choice comes from the least accuracy of low-cost sensors' measurements but limits the size of the datasets used to train and evaluate the prediction models given the lower number of official monitoring stations. An alternative choice would have been to use the low-cost sensors' measurements or part of them as ground truth measurements: this choice leads to larger but noisier datasets.

An other scope for improvement is on how the low-cost sensors' measurements are processed in the prediction model: in the experiments presented here, they are processed in a convolutional neural network along with other variables like humidity and temperature. Given the high impact of this processing on the engine's performance, this part of the engine could be improved further to make the most of low-cost sensors' data.

Finally, this paper focuses on air quality spatial variability. A logical improvement would be to study how low-cost sensors can help to model air quality temporal variability.


\bibliographystyle{ACM-Reference-Format}
\bibliography{air_quality_prediction}










\end{document}